# Electronic Structure of $Ta_2O_5$ Polymorphs: Variation Trend of Band Gap and the Role of Oxygen Vacancies

Hui-Min Tang[a] and Yong Yang[b,a,c]*

[a]*School of Physical Science and Technology, Guangxi Normal University, Guilin 541001, China.*
[b]*Key Lab of Photovoltaic and Energy Conservation Materials, Institute of Solid State Physics, HFIPS, Chinese Academy of Sciences, Hefei 230031, China.*
[c]*Science Island Branch of Graduate School, University of Science and Technology of China, Hefei 230026, China.*

We provide a systematic study on the electronic structure of a series of $Ta_2O_5$ polymorphs using standard density functional theory (DFT) calculations as well as the more accurate many-body perturbation theory within the *GW* approximation. For the defect-free polymorphs, the variation trend of band gap can be microscopically related to the strength of orbital hybridization, and to the macroscopic bulk formation energy. The presence of oxygen vacancies is found to have profound effects on the electronic properties near the Fermi level, notably the reduction of band gap and gap closure (insulator-to-metal transition). Furthermore, depending on the vacancy sites, the band gap of some defective system may be enlarged with comparison to its defect-free counterpart. Such an anomalous behavior originates from the unexpected formation of Ta-Ta bonds.

*Corresponding Author (Y. Y.): yyanglab@issp.ac.cn

## 1. Introduction

Band gap is a crucial physical parameter for characterizing the fundamental properties of materials. It not only distinguishes metals from semiconductors and insulators [1], but also has direct influences on various linear response properties of materials, including thermal conductivity, heat capacity, electrical conductivity, as well as basic optical properties such as absorption, reflection, and refraction [1]. Since the advent of quantum mechanics and its application to calculating electronic energy levels in small molecules and solid-state materials, the determination of energy gap/band gap in polyatomic and condensed matter systems has been a focal point of research efforts [2-11]. When a macroscopic or mesoscopic material undergoes size reduction and eventually transitions to the monomer atoms or molecules of its chemical components, the concept of band gap describing systems with a large number of particles is simplified to the frontier orbitals of simple molecular systems, specifically referring to as the highest occupied molecular orbital (HOMO) and the lowest unoccupied molecular orbital (LUMO) of electrons. These two are essentially identical, indicating the energy difference between the highest occupied state and the lowest unoccupied state. Quantitatively characterizing the evolving trend of band gaps is an inherent pursuit of scientific research. In general, materials with identical chemical compositions but distinct crystal structures exhibit varying band gaps. For instance, graphite and diamond are allotropes of carbon whose crystal structures are different. Graphite has a hexagonal layered structure formed by $sp^2$ hybridization between carbon atoms, while diamond has a face-centered cubic structure through $sp^3$ hybridization of neighboring carbon atoms. Graphite functions as a semi-metal with zero band gap, whereas diamond acts as a wide band gap semiconductor with a band gap of approximately 5.47 eV [12-14].

Naturally, one may ask the following question: For semiconductors/insulators consisting of identical or similar chemical components, is it possible to characterize the variation trend of their band gaps using simply a well-defined quantity (either microscopic or macroscopic)? For simple small molecules consisting of identical atoms such as $H_2$, the variation in energy gap can be semi-quantitatively measured by the strength of overlapping integrals based on linear combination of atomic orbitals (LCAO) [15]. This measurement can be extended to their corresponding molecular crystals which are bound by van der Waals forces. However, it is not straightforward to apply such a description to the band gaps of covalent crystals. Even for monoatomic crystals of covalent type, the variation of band gaps cannot be simply measured by the strength of overlapping integrals between the neighboring atoms alone. A typical example is graphite and diamond as mentioned above, although the distinction of band gaps between the two can be explained through numerical calculations using the tight-binding (TB) model [16, 17] or state-of-the-art *ab initio* methods based on density functional theory (DFT) calculations [11, 18]. On the other hand, for simple molecules composed of dissimilar atoms with large differences in electronegativity, the variation trend of the HOMO-LUMO gap and band gap of their corresponding macroscopic crystals (ionic crystals) can be elucidated by the magnitude of differences in electronegativity between the constituent elements. Typical binary ionic compounds, alkali halides (e.g., NaF, NaCl, and NaBr…) and alkaline earth meal halides (e.g., $CaF_2$, $SrF_2$, $BaF_2$…), for instance, their energy gaps demonstrate a monotonic decrease (increase) with reducing (enhancing) differences in electronegativity [15, 19-22]. Physically, the difference in electronegativity is directly related to the strength of ionic bonds: Larger difference in electronegativity implies stronger attractive forces between the cations and anions and therefore stronger ionic bonds and consequently larger cohesive/formation energies at macroscopic scale. Equivalently, the band gaps of alkali halides (and alkaline earth meal halides) scales linearly with cohesive/formation energies of the systems. This is a well-defined quantity that characterizes the variation trend of band gaps of typical ionic compounds in which the composition elements come from the same main groups but different atomic numbers. Nevertheless, it is unclear whether such a simple quantity is applicable to or not to describe the band gap variations of compound semiconductors/insulators which consist of identical chemical components but with different spatial arrangements (crystal structures).

In this work, we address this topic by studying the band gap variation of transition metal oxide $Ta_2O_5$, one of the most versatile semiconducting materials. $Ta_2O_5$ is widely used in optical coatings [23], high-k dielectrics for high-density transistors [24], corrosion-resistant coatings [25, 26] and catalyst for electro- and photo-catalysis [27–31]. Due to the difference of preparation conditions, many crystalline phases/polymorphs of $Ta_2O_5$ have been proposed theoretically and experimentally. At low temperature, the reported phases are $L_{SR}$ [32], $\beta_R$ [33], $\delta$ [34], $\beta_{AL}$ [35], $T$ [36], $L_{GMR}$ [37]; the $Z$ and $B$ phase of $Ta_2O_5$ are synthesized under high pressure [38, 39]; the $\lambda$ and $\gamma$ phase, which have been theoretically proposed recently [40, 41]; and the very recently identified $\gamma_1$ phase of $Ta_2O_5$, which contains one only formula unit (Z=1) [42]. At high-temperatures ($T \geq 1633$ K) the H phase is found [43, 44]. The reported band gap values for these phases/polymorphs display considerable variations. However, there still lacks a comprehensive physical explanation for the observed trend in both macroscopic and microscopic aspects. Based on state-of-the-art *ab initio* calculations, we have systematically investigated the electronic structure of the low-temperature and ambient-pressure phases of $Ta_2O_5$. Furthermore, we provide a comprehensive elucidation for the variation trend of band gap from both macroscopic and microscopic perspectives. We have observed a strong correlation between the band gap and the material's bulk formation energy at the macroscopic level, which is directly associated with the strength of orbital hybridization among its constituent atoms at the microscopic level. Our work offers a useful scaling factor that characterizes the variation trend of band gaps of stoichiometric metal oxides in a simple and unified manner.

Another factor that has significant impact on the band gap of a semiconductor is structural defects, either foreign or native ones [7-9, 45-50]. Indeed, it is found that by introducing oxygen vacancies into a transition metal oxide such as $Ta_2O_5$ [46-49], both the atomic and electronic structures of the materials are significantly modified, including long-range lattice relaxations as well as band gap reduction. Furthermore, the mobility of oxygen vacancies (referred to as Vo hereafter) in a transition metal oxide plays a central role in the application of redox-based resistive switching memories [46, 50]. Here, we made a systematic study on the electronic properties of oxygen-deficient $Ta_2O_5$ polymorphs. Band gap reduction and insulator-to-metal transition (gap closure) is observed upon the introduction of Vo, even at vacancy concentration of ~ 10%. More interestingly, it is found that depending on the vacancy sites and the local bonding geometries after structural relaxation, the band gap of some defective system may be enlarged with comparison to its defect-free counterpart.

## 2. Computational methods

Our first-principles calculations were carried out using the Vienna *ab initio* simulation package (VASP) [51, 52], which is based on density functional theory (DFT). The electron wave functions are expanded using a plane-wave basis set and the projector-augmented-wave (PAW) potentials are employed for the electron-ion interactions [53, 54]. The exchange-correlation interactions of electrons are described by the generalized gradient approximation (GGA) within the PBE formalism [55]. The energy cutoff for plane waves is 600 eV. In order to ensure the convergence of the calculations, the k-meshes for the calculations of $\gamma$, $\gamma_1$, $\lambda$, $\beta_R$, $L_{SR}$, $\delta$ and $\beta_{AL}$ phase are 10×10×4, 6×6×4, 4×4×8, 6×6×12, 4×2×4, 6×6×12 and 6×12×6, respectively. All the k-meshes are generated by using the Monkhorst-Pack scheme [56].

Since standard DFT-GGA calculations usually underestimate the band gaps of semiconductors and insulators, we have employed state-of-the-art method for band gap calculations, the *GW* method [57, 58], which includes the many-body effects (exchange and correlation) of electrons to calculate the energy levels. We take the one-shot as well as the self-consistent *GW* approach implemented in the VASP code [59] to calculate the energy spectrum. The energy spectra of quasiparticle are obtained by solving the following equation [58]:

$$(T + V_{ext} + V_H)\psi_{nk}(\vec{r}) + \int d\vec{r}\,'\Sigma(\vec{r},\vec{r}';E_{nk})\psi_{nk}(\vec{r}') = E_{nk}\psi_{nk}(\vec{r}), \qquad (1)$$

where $T$ is the kinetic energy operator of electrons, $V_{ext}$ is the external potential due to ions, $V_H$ is the electrostatic Hartree potential, $\Sigma$ is the self-energy operator, and $E_{nk}$ and $\psi_{nk}(\vec{r})$ are respectively the quasiparticle energies and wave functions. Within the *GW* approximation proposed by Hedin [57], $\Sigma$ is calculated as follows:

$$\Sigma(\vec{r},\vec{r}';E) = \frac{i}{2\pi}\int d\omega e^{i\delta\omega} G(\vec{r},\vec{r}';E+\omega) W(\vec{r},\vec{r}';\omega), \qquad (2)$$

where $G$ is the Green's function, $W$ is the dynamically screened Coulomb interaction, and δ is a positive infinitesimal. In the *GW* calculations, the k-meshes of γ, $γ_1$, λ, $β_R$, δ and $β_{AL}$ phase are 6×6×2, 6×6×4, 4×4×8, 4×4×8, 4×4×8 and 4×8×4, respectively, and the number of energy bands involved in the *GW* calculations is 256 in all phases except for the $γ_1$ phase, which is 128.

For both defect-free and defective structures of $Ta_2O_5$, their bulk formation energies ($ΔE_f$) are calculated as follows:

$$\Delta E_f = E[(Ta_2O_{5-x})_n] - nE[Ta_{bulk}] - mE[O_2] \qquad (3)$$

where $n = Z$ is the number of formula units contained in the crystal unit cell, $m = n(5-x)/2$, $x \geq 0$, with $x = 0$ corresponds to the defect-free structure. The terms $E[(Ta_2O_{5-x})_n]$, $E[Ta_{bulk}]$, and $E[O_2]$ are the total energies of $Ta_2O_{5-x}$ system, the bulk Ta metal with a body-center cubic unit cell, and an isolated $O_2$ molecule at its spin-triplet state. The introduction of oxygen vacancies involves the breaking of Ta-O bonds and consumption of energies, which may be evaluated using vacancy formation energy as defined below:

$$E_{vf} = E[(Ta_2O_{5-x})_n] + \left(\frac{xn}{2}\right)E[O_2] - E[(Ta_2O_5)_n] \qquad (4)$$

To gain a deeper understanding of the possible formation of Ta-Ta bonds around the defective sites from the level of electron, we have calculated the charge density difference of oxygen-deficient $Ta_2O_5$ system as follows:

$$\Delta \rho = \rho[(Ta_2O_{5-x})_n] - \rho[Ta] - \rho[(Ta_{2-y}O_{5-x})_n] \qquad (5)$$

where $\rho[(Ta_2O_{5-x})_n]$ is the charge density of the $Ta_2O_{5-x}$ system, $\rho[Ta]$ represents the charge density of a single Ta atom, and $\rho[(Ta_{2-y}O_{5-x})_n]$ is the charge density of the $Ta_2O_{5-x}$ system after removing one Ta atom, with $y = 1/n$.

## 3. Results and discussion

We begin with investigations on the electronic structure of defect-free $Ta_2O_5$ polymorphs; then the oxygen-deficient systems are considered. Special attention is paid to the energy levels near the Fermi level, which determine the band gap value and the fundamental excitation properties of a given system.

### A) Electronic properties of defect-free $Ta_2O_5$

The electronic band structures and density of states (DOS) of six low-temperature $Ta_2O_5$ phases which share some common features of layer packing in the atomic scale [42], are presented in Fig. 1. The data are calculated using both DFT-GGA and $GW$ method. To our knowledge, this is the first comprehensive study on the accurate band gaps of $Ta_2O_5$ polymorphs. Meanwhile, the band gaps are listed in Table 1, together with the top of the valence band (VBM), and the bottom of the conduction band (CBM).

**Table 1.** The values of VBM, CBM and band gap $E_g$ (in units of eV) of the seven $Ta_2O_5$ phases calculated by DFT-GGA, one-shot ($G_0W_0$) and self-consistent $GW$ method. Due to the extremely large computational burden of $GW$ calculations, only GGA data for the $L_{SR}$ phase are available. The formation energies ($\Delta E_f$) by DFT-GGA calculations are also presented.

| phase | γ | γ₁ | λ | $L_{SR}$ | $β_R$ | δ | $β_{AL}$ |
|---|---|---|---|---|---|---|---|
| VBM | 0.83 | 0.93 | 2.02 | 1.91 | 1.87 | 2.00 | 2.33 |
| CBM | 3.13 | 3.18 | 4.10 | 3.90 | 3.79 | 3.10 | 2.54 |
| $E_g^{GGA}$ | 2.30 | 2.25 | 2.08 | 1.99 | 1.92 | 1.10 | 0.21 |
| VBM | 0.80 | 0.89 | 2.04 | | 1.87 | 2.06 | 2.33 |
| CBM | 4.16 | 4.24 | 5.11 | | 4.75 | 3.87 | 3.30 |
| $E_g^{G0W0}$ | 3.36 | 3.35 | 3.07 | | 2.88 | 1.81 | 0.97 |
| VBM | 0.50 | 0.57 | 1.78 | | 1.60 | 1.82 | 2.09 |
| CBM | 4.55 | 4.65 | 5.53 | | 5.10 | 4.28 | 3.70 |
| $E_g^{GW}$ | 4.05 | 4.08 | 3.75 | | 3.50 | 2.46 | 1.61 |
| $\Delta E_f$ (eV/Ta) | -10.038 | -9.993 | -9.904 | -9.871 | -9.835 | -9.043 | -8.863 |

The γ phase displays the largest band gap, which is indirect with a value of ~ 2.30 eV by GGA. The top of the valence band (VBM) is located at the M point, while the bottom of the conduction band (CBM) is situated at the Γ point. However, this value by DFT-GGA calculation is still notably smaller than the value given by experimental measurement, which is approximately 4 eV [60, 61], and would be amended by the self-consistent *GW* calculations as discussed below. The band gap of $\beta_{AL}$ phase is determined to be the smallest, with a value of 0.21 eV, as obtained from DFT-GGA calculations. The majority of the observed band gaps in these seven phases (including $L_{SR}$, see Table 2) are indirect, with the exception of $\lambda$ and $\beta_{AL}$, as indicated in Table 2. The phase $\lambda$ exhibits a direct band gap of 2.08 eV at the Z point, while $\beta_{AL}$ also displays a direct band gap of 0.21 eV which is located at the Y point.

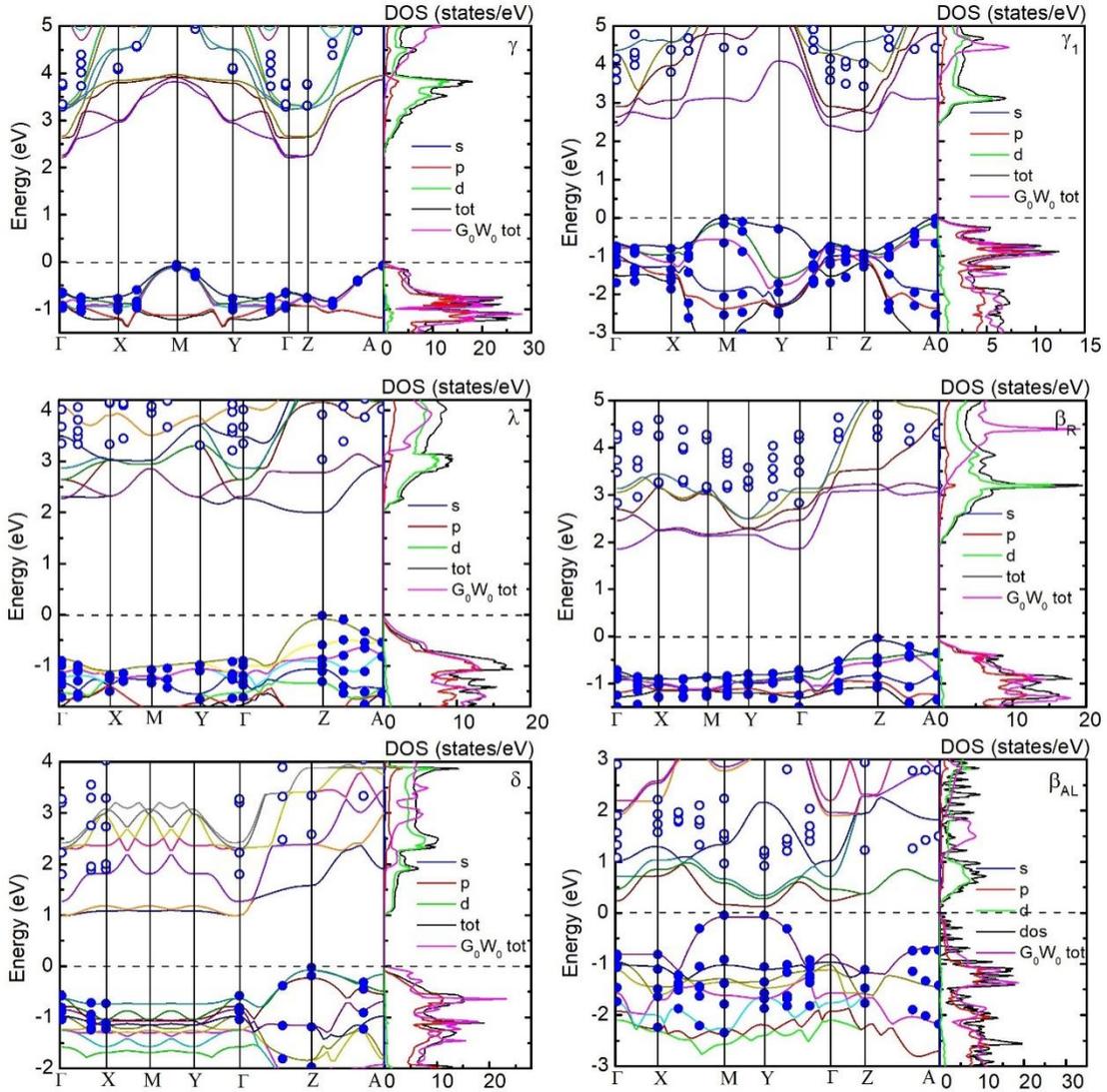

**Fig. 1**. The band structures, density of states (DOS), and partial density of states (PDOS) for six phases ($\gamma$, $\gamma_1$, $\lambda$, $\beta_R$, $\delta$, $\beta_{AL}$) of $Ta_2O_5$ calculated using DFT-GGA and GW methods. In the band diagram, the solid line represents the band results along highly symmetric k-point lines obtained from GGA calculations, while the solid and hollow blue circles represent results obtained from GW calculations. The Fermi level (indicated by vertical dashed lines) is set at 0 eV. This convention applies to all the figures.

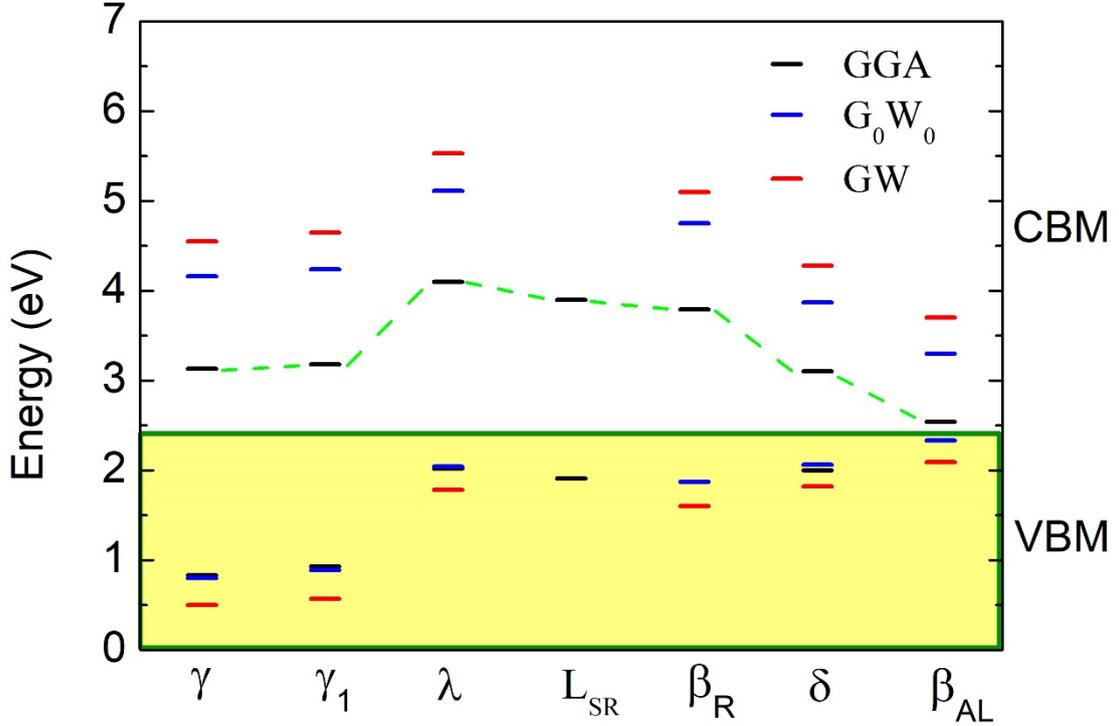

**Fig. 2.** Schematic diagrams for the level positions of the VBM and CBM determined using GGA, one-shot GW ($G_0W_0$), and self-consistent GW methods.

The partial density of states (PDOS) reveals that the electronic states near the top of the valence bands of all phases are predominantly contributed by the O $2p$ orbitals, while the electronic states near the bottom of the conduction bands are dominated by the Ta $5d$ orbitals.

    As compared to GGA type calculations, the self-energy corrections in the *GW* method significantly improve the band gaps. Figure 2 schematically illustrates the positions of the VBM and CBM obtained using GGA, one-shot GW ($G_0W_0$) and self-consistent GW methods (GW). The indirect band gap of the $\gamma$ phase predicted by one-shot and self-consistent GW calculations increases to 3.36 eV and 4.05 eV, respectively, with the latter comparing well with experimental data [60, 61]. The VBM obtained by GGA and GW are found to exhibit negligible differences, and the

corrections to band gap are primarily attributed to variations of CBM [11]: $E_{CBM} \approx \varepsilon_{CBM} + \langle \psi_{CBM}|\Sigma(E_n) - V_{xc}|\psi_{CBM}\rangle$ , where the term ( $\Sigma(E_n) - V_{xc}$ ) represents self-energy corrections to the exchange-correlation interactions within the *GW* approximation. For all the phases, it is clearly seen that the CBM values calculated using *GW* exhibit significant increase compared to those derived from GGA, thereby resulting in enhanced band gaps.

**Table 2.** Calculated of *lm*-components of the wave functions of the VBM and CBM of the seven $Ta_2O_5$ phases. The descriptor for the strength of orbital hybridization $f_\lambda = e^{-\lambda}$, as defined by Eq. (6).

| phase | | s | $p_y$ | $p_z$ | $p_x$ | $d_{xy}$ | $d_{yz}$ | $d_{z^2}$ | $d_{xz}$ | $d_{x^2-y^2}$ | tot | $f_\lambda$ |
|---|---|---|---|---|---|---|---|---|---|---|---|---|
| γ | VBM (M) | 0.000 | 0.330 | 0.091 | 0.352 | 0.006 | 0.001 | 0.000 | 0.001 | 0.000 | 0.781 | 0.720 |
| | CBM (Γ) | 0.000 | 0.005 | 0.000 | 0.004 | 0.867 | 0.018 | 0.002 | 0.005 | 0.007 | 0.906 | 0.678 |
| $γ_1$ | VBM (A) | 0.000 | 0.394 | 0.151 | 0.235 | 0.002 | 0.002 | 0.000 | 0.001 | 0.000 | 0.784 | 0.747 |
| | CBM (Z) | 0.000 | 0.000 | 0.000 | 0.016 | 0.814 | 0.000 | 0.000 | 0.069 | 0.000 | 0.899 | 0.685 |
| λ | VBM (Z) | 0.000 | 0.134 | 0.395 | 0.263 | 0.000 | 0.000 | 0.000 | 0.002 | 0.000 | 0.793 | 0.723 |
| | CBM (Z) | 0.035 | 0.024 | 0.001 | 0.013 | 0.216 | 0.000 | 0.048 | 0.000 | 0.561 | 0.898 | 0.666 |
| $L_{SR}$ | VBM (Z) | 0.000 | 0.082 | 0.411 | 0.289 | 0.000 | 0.001 | 0.000 | 0.003 | 0.000 | 0.787 | 0.674 |
| | CBM (Γ) | 0.000 | 0.000 | 0.042 | 0.000 | 0.000 | 0.583 | 0.000 | 0.287 | 0.000 | 0.912 | 0.718 |
| $β_R$ | VBM (Z) | 0.000 | 0.279 | 0.381 | 0.126 | 0.000 | 0.001 | 0.000 | 0.001 | 0.000 | 0.789 | 0.690 |
| | CBM (Γ) | 0.000 | 0.000 | 0.003 | 0.000 | 0.000 | 0.226 | 0.000 | 0.681 | 0.000 | 0.910 | 0.610 |
| δ | VBM (Z) | 0.000 | 0.187 | 0.357 | 0.232 | 0.000 | 0.004 | 0.000 | 0.005 | 0.000 | 0.785 | 0.624 |
| | CBM (Γ) | 0.000 | 0.000 | 0.022 | 0.000 | 0.000 | 0.662 | 0.000 | 0.245 | 0.000 | 0.929 | 0.447 |
| $β_{AL}$ | VBM (Y) | 0.000 | 0.467 | 0.322 | 0.000 | 0.000 | 0.000 | 0.000 | 0.000 | 0.000 | 0.789 | 0.597 |
| | CBM (Y) | 0.000 | 0.000 | 0.000 | 0.000 | 0.000 | 0.000 | 0.000 | 0.906 | 0.000 | 0.906 | 0.553 |

Table 1 presents the band gaps in descending order as follows: $γ \approx γ_1 > λ > L_{SR} \approx β_R > δ > β_{AL}$. From their bulk formation energies (Table 1), the stability of these phases follows a descending order from the largest to the smallest with $γ \approx γ_1 > λ > L_{SR} \approx β_R > δ > β_{AL}$. The variation trend of band gap and the phase stability, as indicated by the magnitude of bulk formation energies, exhibits an obvious one-to-one correspondence, implying a potential correlation between them.

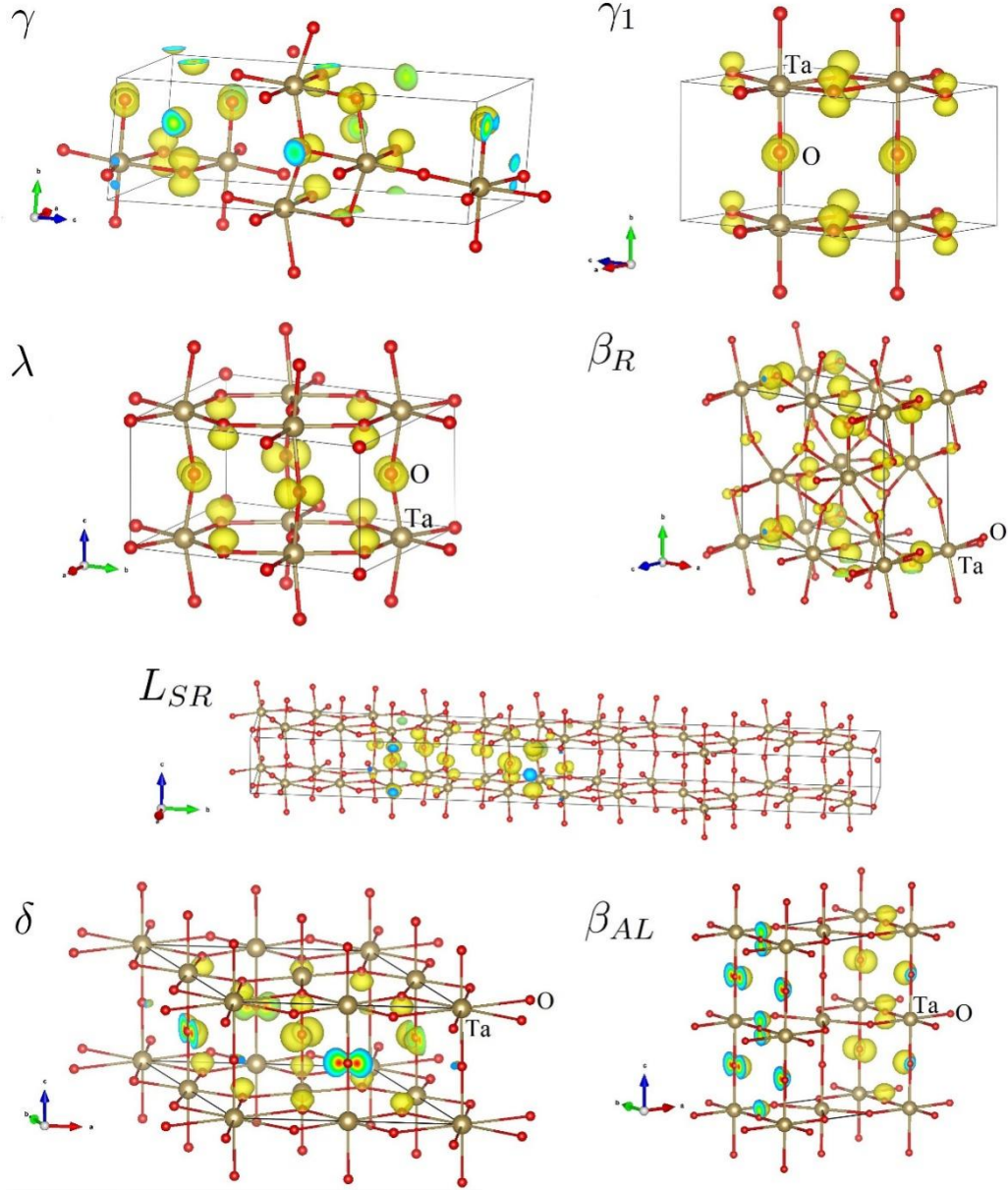

**Fig. 3.** Band-decomposed charge density of VBM for the seven phases of $Ta_2O_5$.

Indeed, the variation trend of bulk formation energies and band gaps coincide with each other. Considering the large difference between the electronegativities of O and Ta, such a consistency leads one to conjecture that the variation of band gaps of $Ta_2O_5$ polymorphs may be similar to that of typical ionic compounds [15, 19-22]. It should be noted here that the nearly identical band gap and formation energy of $\beta_R$ and $L_{SR}$ arise from the fact that these two polymorphs essentially share the same basic building blocks (i.e., $TaO_6$ and $TaO_7$) [33]. Similar situation is found in the $\gamma$ and $\gamma_1$ phase [42]. Physically, the formation energy correlates directly with the strength of the Ta-O bond. From a microscopic perspective, the strength of this chemical bond is intricately linked to the degree/strength of orbital hybridization. Therefore, we have further examined the orbital characteristics of VBM of each $Ta_2O_5$ phase, which contains typical orbital hybridization features of the bonding states. The results are presented in Table 2.

It is clearly seen from Fig. 3 that the VBM of $Ta_2O_5$ are primarily composed of O $2p$ orbitals. Importantly, there are significant variations in the charge distribution of the VBM among these $Ta_2O_5$ phases. The charges of the VBM uniformly distribute across almost all O atoms in the primitive cell of phases that are both energetically and dynamically stable. For example, the $\gamma$ and $\gamma_1$ phases as shown in Fig. 3, demonstrate a nearly uniform charge distribution on each O atom. Conversely, in the less stable phases, the charge of the VBM is predominantly localized on a small number of O atoms. The $\delta$ and $\beta_{AL}$ phase, for instance, the electrons at VBM primarily localize on four O atoms. It is therefore evident that the stability of the phases exhibits a close correlation with the degree/strength of orbital hybridization, which may be described using the overlap integral within the framework of LCAO (see Appendix). For transition metal oxides such as $Ta_2O_5$, the chemical bonds formed between metal and O atoms generally show both ionic and nontrivial covalent characteristics. The bond strength is largely determined by the energy gain due to the overlap of constituent atomic wave functions, which may be described by the sum of overlap integrals (Appendix). Therefore, we propose the following scaling factor to quantitatively assess the degree/strength of orbital hybridization:

$$f_\lambda = e^{-\lambda} \tag{6}$$

where $\lambda = \sum_{n=1}^{9} \Delta_n$, $\Delta_n = \sqrt{\sum_{j=1}^{N}(|c_{nj}|^2 - \frac{|c_n|^2}{N})^2}$, and $N$ is the total number of atoms, and the number $n$ denotes the nine $lm$-resolved orbitals ($s$, $p_y$, $p_z$, $p_x$, $d_{xy}$, $d_{yz}$, $d_{z2}$, $d_{xz}$, $d_{x2-y2}$). The coefficients satisfy $\sum_{j=1}^{N}|c_{nj}|^2 = |c_n|^2$, which are readily obtained from DFT calculations. According to the definition, the maximum orbital hybridization is obtained when each orbital and atom contributes uniformly to the wave function of the energy levels. This corresponds to the possible largest overlap of atomic wave functions (Appendix). Obviously, at the ground state one has $f_\lambda = 1$. For the extreme situation when the wave function is localized on a single atom and a single orbital, $f_\lambda = e^{-(1-\frac{1}{N})}$. The computed values of $f_\lambda$ for each Ta$_2$O$_5$ phase are listed in Table 2. It is seen from Table 2 that in the γ phase, which possesses the largest band gap, the wave function is nearly uniformly distributed across each orbital and atom in the unit cell, resulting in $f_\lambda = 0.720$ and $0.678$ for the VBM and CBM, respectively. By contrast, for the $β_{AL}$ phase with the smallest band gap, the wave function of VBM is largely localized with the characteristics of $p_y$ and $p_z$ orbitals while CBM is confined to a single $d_{xz}$ orbital. Consequently, this leads to a smaller $f_\lambda = 0.597$ and $0.593$ for the VBM and CBM, respectively.

The band gaps of the seven Ta$_2$O$_5$ polymorphs, as calculated using GGA, G$_0$W$_0$, and GW methods, are depicted in Fig. 4. Figure 4(a) illustrates the correlation between band gap and the macroscopically measurable quantity -- the bulk formation energy (in form of its absolute value), while Fig. 4(b) demonstrates the relationship between band gap and the microscopic scaling factor $f_\lambda$. As seen from Fig. 4(a), among these seven phases, structures with higher formation energy exhibit greater stability and consequently larger band gaps. A linear fitting analysis reveals the correlation between band gap and the bulk formation energy ($|ΔE_f|$):

$$E_g = 1.833|ΔE_f| + X, \tag{7}$$

where $X$= -15.999, -15.050, and -14.382 correspond to the band gaps calculated using GGA, $G_0W_0$, and the self-consistent $GW$ methods, respectively. At the macroscopic level, this formula provides an approximate description for the variation trend of band gap across different Ta$_2$O$_5$ polymorphs. At the microscopic level, Fig. 4(b) illustrates the trend of band gap variations with the scaling factor $f_\lambda$. The larger the uniformity of the wave function distribution is in the structure, the higher $f_\lambda$ and subsequently larger band gap is observed. Linear regression analysis reveals a quantitative relationship between band gap and $f_\lambda$ as follows:

$$E_g = 16.473 f_\lambda + Y, \tag{8}$$

where $Y$ = -9.544, -8.686, and -8.017 for the band gaps calculated using GGA, $G_0W_0$, and self-consistent $GW$ methods, respectively. It can be seen that the bulk formation energy of $Ta_2O_5$ and the bond strength offer a better description of the variation trend of band gaps. On the other hand, the microscopic scaling factor $f_\lambda$ provides a barely satisfactory description for the trend in band gap variation among different $Ta_2O_5$ phases. This is due to the fact that the overlap integral (see appendix) itself (consequently the scaling factor) does not take into account the exchange-correlation interactions of electrons, which are important for accurate evaluation of bulk formation energies of metal oxides.

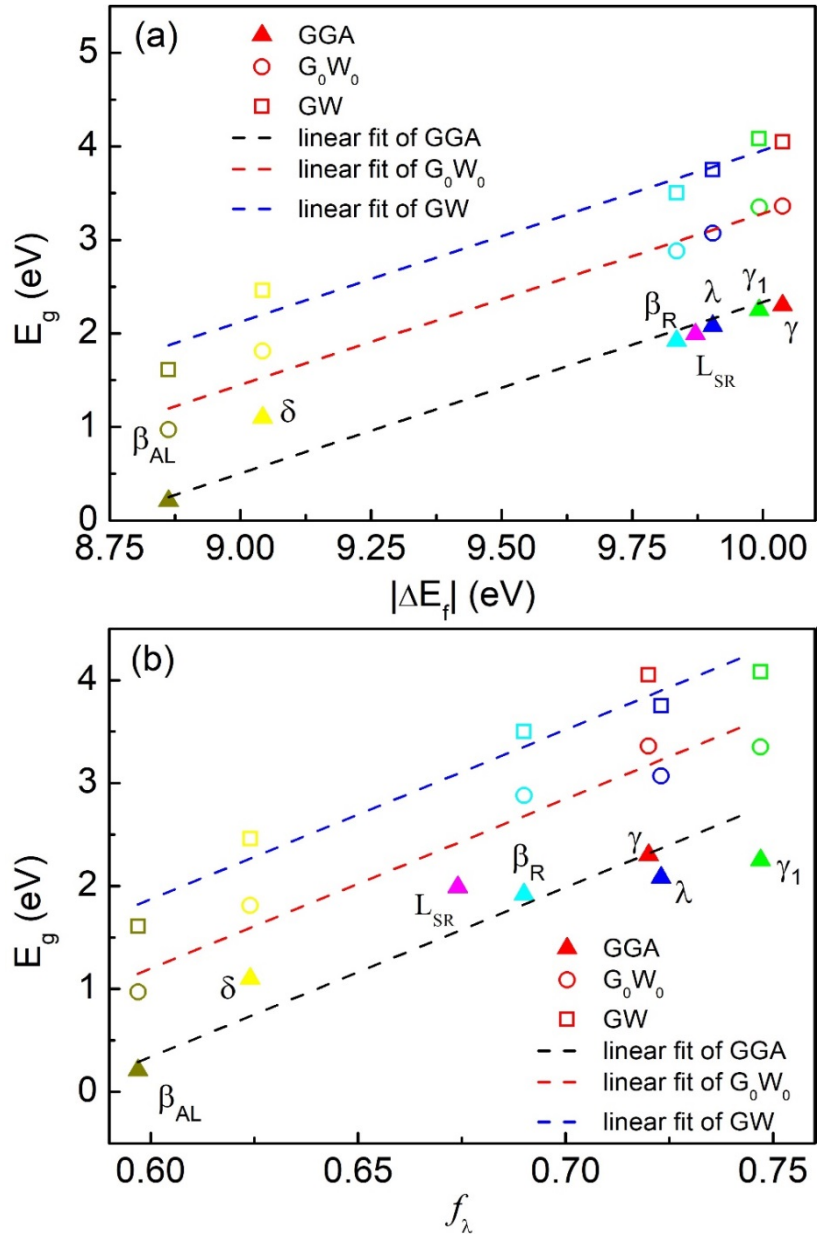

**Fig. 4.** The dependence of band gap and bulk formation energy $|\Delta E_f|$ (a), and the scaling factor $f_\lambda$ of VBM (b) that characterizes the strength of orbital hybridization.

### B) Ta$_2$O$_5$ with oxygen vacancies

In this subsection, we study the effects of oxygen vacancy (Vo) on the electronic properties of Ta$_2$O$_5$. In particular, we shall show how the band gap may be closed by introducing Vo at certain atomic sites. From their bonding geometries with Ta, the O atoms in Ta$_2$O$_5$ may be classified into two types: The O bonded with two Ta (two-coordination) and the O bonded with three Ta (three-coordination). Accordingly, we have two types of oxygen vacancies. For clarity, the Vo created by removing two-coordination O from bulk Ta$_2$O$_5$ is referred to as type 1, and the Vo created by removing three-coordination O is labeled as type 2.

**Table 3.** Calculated formation energies of bulk phase ($\Delta E_f$) and that for the creation of oxygen vacancies ($E_{vf}$) with a content of 10%, and the resulted band gap ($E_g$) of a number of oxygen-deficient Ta$_2$O$_5$ systems. The data for type 1 (two-coordination) and type 2 (three-coordination) oxygen vacancies are labeled by the subscript 1 and 2, respectively.

| Ta$_2$O$_{5-x}$ | $\Delta E_{f1}$(eV/Ta) | $E_{vf1}$ (eV) | $E_{g1}$ (eV) | $\Delta E_{f2}$(eV/Ta) | $E_{vf2}$ (eV) | $E_{g2}$ (eV) |
|---|---|---|---|---|---|---|
| γ | -8.739 | 5.192 | 0 | -8.806 | 4.926 | 0 |
| γ$_1$ | -8.552 | 5.767 | 0.237 | -8.648 | 5.381 | 1.109 |
| λ | -8.571 | 5.331 | 0 | -8.728 | 4.702 | 1.382 |
| β$_R$ | -8.508 | 5.307 | 0 | -8.575 | 5.042 | 0.026 |
| δ | -7.921 | 4.488 | 0 | -8.575 | 1.875 | 0.134 |
| β$_{AL}$ | -8.020 | 3.371 | 0 | -8.517 | 1.380 | 1.975 |

Figure 5 shows the calculated electron DOS at a vacancy concentration of 10%, for the Ta$_2$O$_5$ polymorphs studied in Part A. Such a concentration corresponds to removal of one O from the unit cell of Z=2 polymorphs. For the Z=1 phase (γ$_1$), a (1×1×2) supercell is employed for the simulation of Vo. For all the Z=2 polymorphs (γ, λ, β$_R$, δ, β$_{AL}$), gap closure ($E_g = 0$) and therefore metallization are observed upon the creation of type 1 Vo. Table 3 presents the band gaps and the energy parameters related to the formation of the oxygen-deficient $Ta_2O_5$ polymorphs with type 1 and type 2 Vo. As seen from Fig. 5 and Table 3, both types of Vo lead to the metallization of γ phase, which is energetically the most stable among all the known $Ta_2O_5$ phases. Moreover, it is also found from Table 3 that for the defective structures ($Ta_2O_{5-x}$) which are created from the same defect-free $Ta_2O_5$ phase, a larger bulk formation energy ($|\Delta E_f|$) is associated with a larger band gap ($E_g$). That is, to the defective structure of a given $Ta_2O_5$ polymorph, $|\Delta E_{f2}| \geq |\Delta E_{f1}|$ yields $E_{g2} \geq E_{g1}$, and vice versa. This is in line with the results presented in Part A where strong Ta-O bonds correspond to large band gaps.

Compared to defect-free $Ta_2O_5$, the presence of Vo leads to dangling Ta-O bonds and unpaired electrons in the defective structures and causes large modifications on the electronic states near the Fermi level. Figure 6 displays the *s*, *p*, *d*-resolved PDOS of δ- and $β_{AL}$-$Ta_2O_5$ in the presence of type 1 and type 2 Vo. It is obvious that 5*d* orbitals from Ta atoms dominate the states near the valence and conduction bands, in direct contrast to the defect-free structures where the valence and conduction bands are contributed by 2*p* and 5*d* orbitals (Fig. 1), respectively.

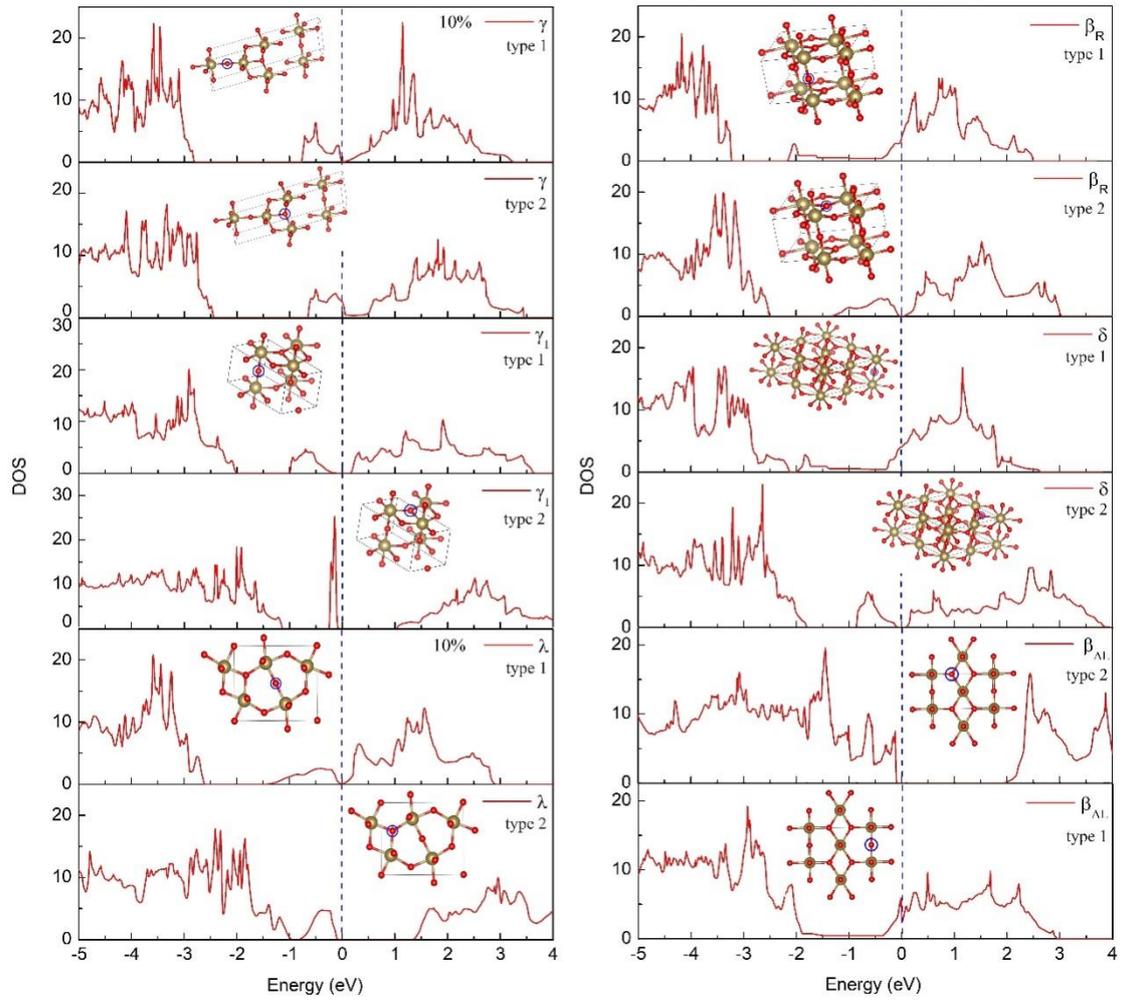

**Fig. 5.** Calculated electron density of states (DOS) for $Ta_2O_5$ polymorphs at a vacancy concentration of 10%.

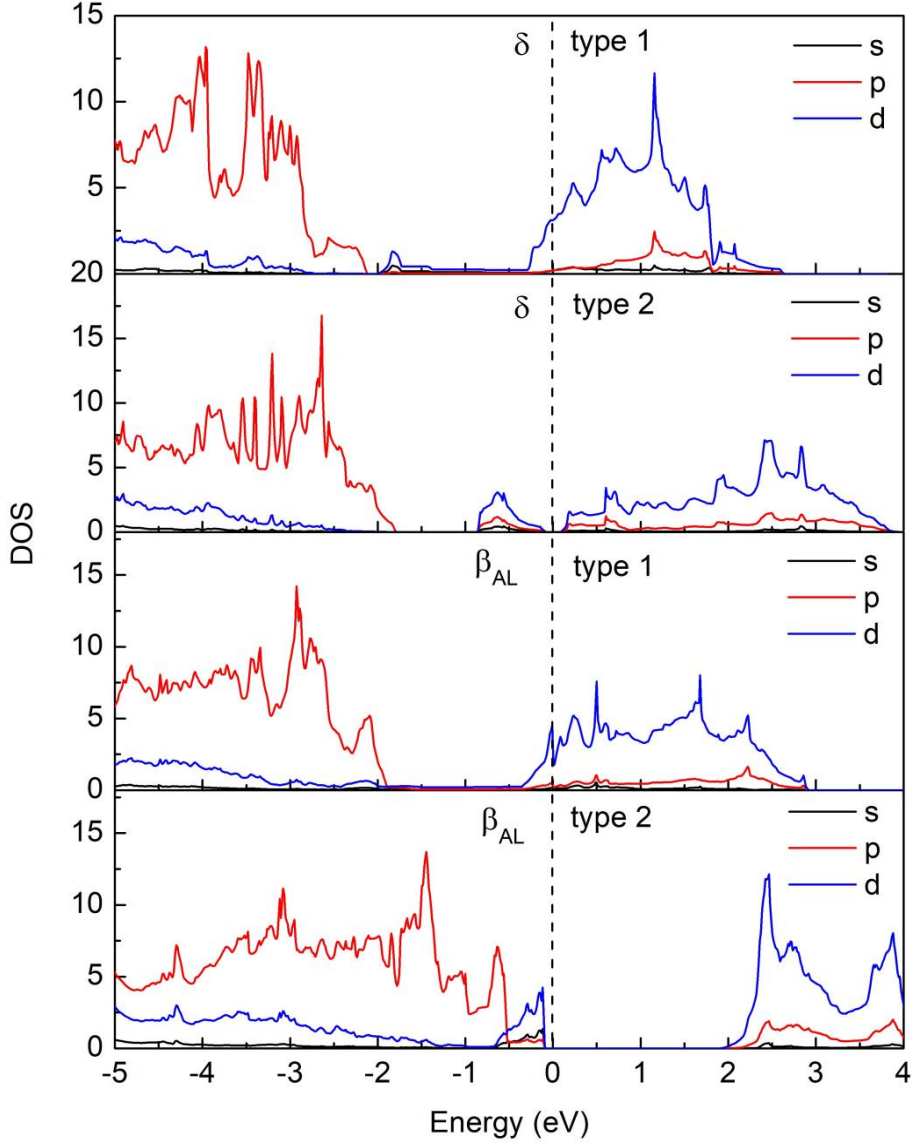

**Fig. 6.** Calculated PDOS of defective δ and $β_{AL}$ $Ta_2O_5$ at a Vo concentration of 10%, for type 1 and type 2 Vo.

In addition to the observation of gap closure in all the Z=2 phases upon the creation of type 1 Vo, another striking feature is the sharp contrast of electronic structure induced by the two types of Vo in $β_{AL}$ phase (Fig. 5): Type 1 Vo leads to gap closure and metallization while type 2 Vo opens up an even larger band gap with comparison to the defect-free structure (1.975 eV vs 0.2 eV by GGA). To reveal the role of dangling bonds in the gap closure/opening process, we plot in Fig. 7 the atom-resolved DOS due to the four Ta atoms in the defective and defect-free $β_{AL}$ phase. In the defect-free $β_{AL}$ phase of $Ta_2O_5$, the contribution of a single Ta atoms (labeled by integers 1, 2, 3, 4) to the DOS near the bottom of conduction bands differ significantly: Compared to the third (Ta3) and fourth (Ta4) Ta atom, the first (Ta1) and second (Ta2) Ta contributed negligibly small to the DOS. In this case, the Fermi level ($E_F$) lies near the bottom of conduction bands (Fig. 7(a)). When one of the two-coordination O bonded with Ta1 (and its periodic image) is removed and type 1 Vo is created, the $5d$ electrons from the dangling Ta-O bond enable Ta1 to have nontrivial contribution to the electronic states near the bottom of conduction bands (Fig. 7(b)). Due to the loosely bonded $5d$ electrons in the dangling bond of Ta1, the Fermi level is shifted upwards (Fig. 7(b)) to the shoulder of the DOS peak near the conduction bottom. On the other hand, when one of the three-coordination O bonded with Ta1, Ta3 (and its periodic image) is removed and type 2 Vo is created, the Fermi level is significantly shifted down to top of valence band and the band gap is enlarged (Fig. 7(c)). Within the rigid band model, the positional shift of the Fermi level upon the creation of Vo explains the gap closure/opening-up in $\beta_{AL}$ $Ta_2O_5$.

Then one may wonder in case of type 2 Vo, how do the $5d$ electrons in the dangling bonds find their positions? The enlarged band gap and the significant drop of the Fermi level imply the possibility of new bond formation. Indeed, the creation of type 2 Vo reduces the Ta1-Ta3 separation from the original value of ~ 3.74 Å to ~ 3.03 Å, a distance comparable to the Ta-Ta bond length (~ 2.86 Å) in a body-centered cubic Ta metal. This is a clear indication of the formation of Ta-Ta bond. To demonstrate this point, we show in Fig. 8 the charge density distribution of VBM ($|\psi_{VBM}|^2$) for the defect-free and defective $\beta_{AL}$ with type 1 & type 2 Vo. Compared to the $2p$ orbitals of VBM in defect-free structure (Fig. 8(a)), the VBM of defective structure with type 1 Vo exhibits clear characteristics of $5d$ orbitals ($d_{xz}$) which is localized abound a single Ta atom (Fig. 8(b)). By contrast, a large portion of electrons of VBM are shared by two neighboring Ta atoms (Ta1 and Ta3, see Fig. 8(c)). Such a charge distribution between two neighboring Ta atoms differs significantly from that in a defect-free structure, and is a clear evidence of Ta-Ta bond. Additionally, we have further analyzed the charge density differences of the two types of Vo, as shown in Fig. 9. Specifically, for type 1 Vo, as depicted in Figs. 9 (a) and 9(b), no charge transfer occurs between two neighboring Ta atoms. By contrast, Figs. 9 (c) and 9(d) unequivocally demonstrate that charge transfer occurs between the two neighboring Ta atoms, providing compelling evidence of the formation of Ta-Ta bond. The unexpected formation of Ta-Ta bond near the vacancy site results in extra energy gain, decrease of Fermi level, splitting of $5d$ orbitals (corresponds to the bonding and anti-bonding states, respectively) and an enlarged band gap in defective $\beta_{AL}$ phase.

To further study the effects of oxygen vacancy on the distribution of charge densities, we plot in Fig. 8(d) the Bader charge [62, 63] associated with the four Ta atoms in the defect-free and defective $\beta_{AL}$ $Ta_2O_5$. In the defect-free structure, Ta1 and Ta2 have the same Bader charge ($\sim 0.072e$) while Ta3 and Ta4 have the same Bader charge ($\sim 0.773e$). Large variations are induced by the presence of Vo. For type 1 Vo, the largest variation in Bader charge comes from Ta1, which was bonded with the removed O at the vacancy site. For type 2 Vo, the Bader charges of Ta1, Ta2, Ta3 increase significantly while that of Ta4 decreases slightly. The rebalance of charge distribution among the Ta atoms leads to significant modifications of electronic properties as shown above. Compared to the type 1 Vo induced Bader charge redistribution around Ta atoms, it is found that much more uniform charge distribution is induced by type 2 Vo. Such a charge distribution provides mutual corroboration for the microscopic picture that the uniformity of charge distribution and the band gap are positively correlated.

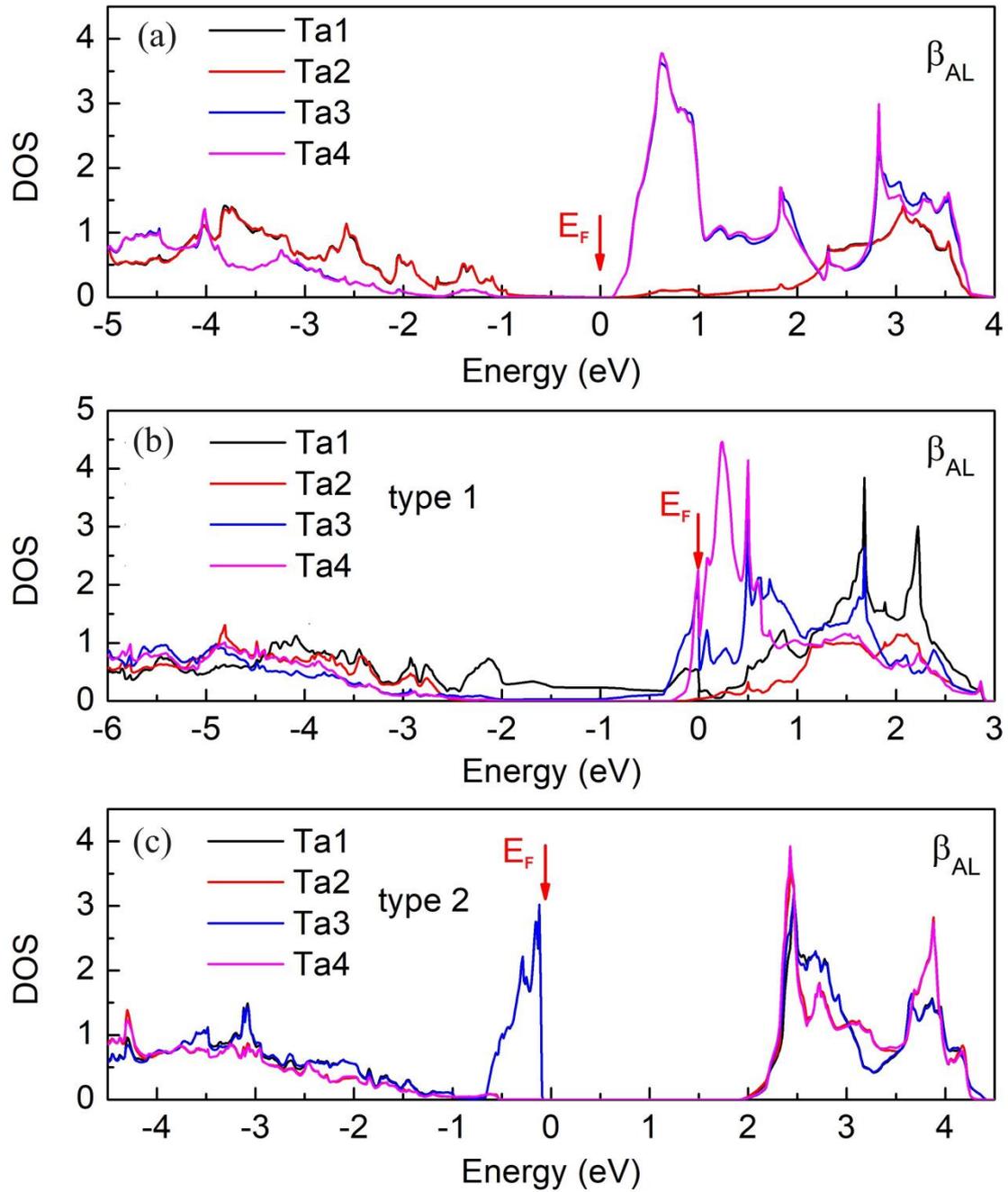

**Fig. 7.** Total DOS contributed by the four Ta atoms in defect-free (a), type 1 Vo (b), and type 2 Vo (c) of $\beta_{AL}$-$Ta_2O_5$. The position of Fermi level is indicated by vertical arrows.

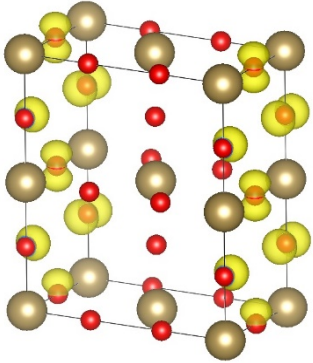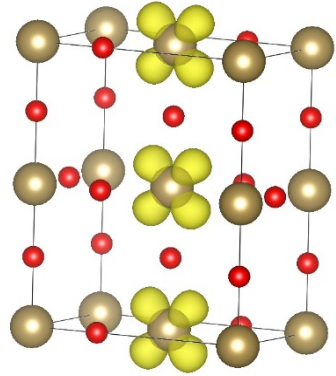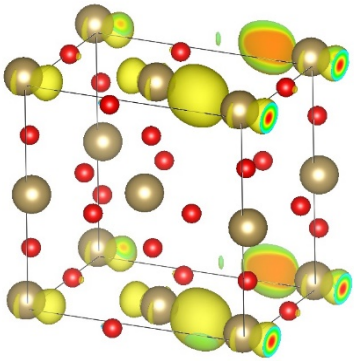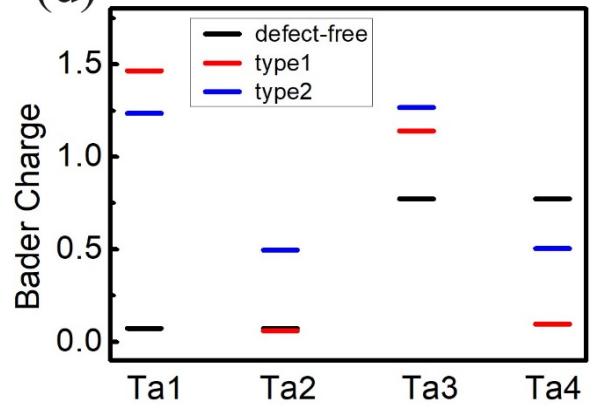

**Fig. 8.** Panels (a)-(c): Spatial distribution of the electrons of VBM in the defect-free (a), type 1 Vo (b), and type 2 Vo (c) of $\beta_{AL}$-$Ta_2O_5$. Panel (d): The Bader charge (in units of elementary electric charge $e$) associated with the four Ta atoms for the defect-free and defective structures of $\beta_{AL}$-$Ta_2O_5$.

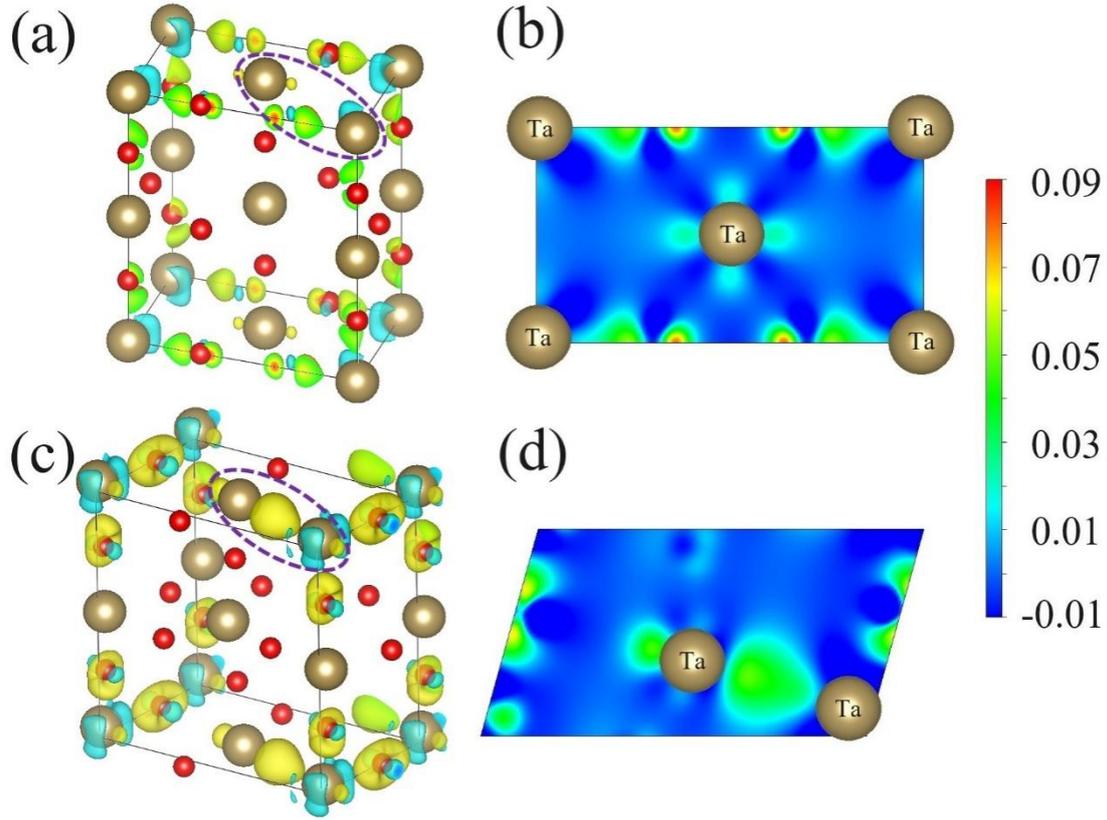

**Fig. 9.** Charge density differences of the type 1 Vo (a), and type 2 Vo (c) of $\beta_{AL}$-$Ta_2O_5$. Panels (b) and (d) are the 2D contours of the charge densities, which are sliced along the surface normal plane through the centers of the two circled Ta atoms as highlighted in (a) and (c).

## 4. Conclusions

Based on first-principles calculations, we have conducted a comprehensive and systematic investigation into the electronic properties of a number of $Ta_2O_5$ polymorphs, with focus on the variations of band gaps in defect-free and defective structures. The variation of band gaps is found to be directly correlated with the bulk formation energy at the macroscopic level and the strength of orbital hybridization at the microscopic level. In general, crystals with higher stability are endowed with larger band gaps, which are related to higher uniformity of the spatial distribution of the bonding states. Based on the data obtained from GGA and GW calculations, empirical formulas are suggested for quantitatively relating the variation of band gap to bulk formation energy and to the strength of orbital hybridization. With the introduction of oxygen vacancies, the electronic structures of $Ta_2O_5$ polymorphs are significantly modified. Band gap reduction and closure are observed in most of the defective systems. Depending on the vacancy sites, the impact on band gap induced by oxygen vacancies may differ a lot. A typical example is defective $β_{AL}$-$Ta_2O_5$: Type 1 Vo leads to gap closure and insulator-to-metal transition while type 2 Vo causes an enhanced band gap with comparison to the defect-free structure. Such an anomalous behavior originates from the unexpected formation of Ta-Ta bonds in the vacancy sites. In brief, our work provides a simple way for characterizing the variation trend of band gaps of stoichiometric metal oxides, and demonstrates the feasibility of qualitatively measuring the band gaps of metal oxides simply based on their formation energies. The method employed here may be extended to the study of the other wide-gap transition metal oxides which comprise of multiple structural phases.

## Acknowledgements

This work is financially supported by the National Natural Science Foundation of China (No. 12074382, 11804062, 11474285). We are grateful to the staff of Hefei Branch of Supercomputing Center of Chinese Academy of Sciences, and Hefei Advanced Computing Center for their support of supercomputing facilities.

**Appendix**

In this appendix, we derive the condition at which the overlap integral of two types (labeled as A and B) of constituent atoms has its maximum. In a polyatomic and/or condensed matter system, the electron wave function may be expanded using a complete basis sets (hybridization of atomic orbitals) as: $\Psi = \sum_n c_n \psi_n$. Using the linear combination of atomic orbitals (LCAO), the basis $\psi_n$ can be expressed as the sum product of a radial part $\chi(r)$ and the spherical harmonics $Y_{lm}(\theta, \varphi) \equiv Y_n(\theta, \varphi)$:

$$\psi_n = \sum_{k=1}^{N} g_k \chi_{nk}(r) Y_n(\theta, \varphi)$$
$$= \sum_{i=1}^{n_A} g_{Ai} \chi_{nAi}(r) Y_n(\theta, \varphi) + \sum_{j=1}^{n_B} g_{Bj} \chi_{nBj}(r) Y_n(\theta, \varphi), \quad (A1)$$

where the subscript $n$ labels the $n$th spherical harmonics under the classification of angular and magnetic quantum numbers $l$ and $m$; the subscript index $i$ and $j$ denotes the atomic sites; $n_A$, $n_B$ is the number of A and B atoms, respectively; $n_A + n_B = N$ is total number of atoms. The expanding coefficients are subjected to the constraint $\sum_n |c_n|^2 = 1$, $\sum_j |g_j|^2 = 1$. Let $\phi_{nAi} = \chi_{nAi}(r) Y_n(\theta, \varphi)$, $\phi_{nBj} = \chi_{nBj}(r) Y_n(\theta, \varphi)$, then

$$\psi_n = \sum_{i=1}^{n_A} g_{Ai} \phi_{nAi} + \sum_{j=1}^{n_B} g_{Bj} \phi_{nBj} \quad (A2)$$

Within the independent electron approximation, the Hamiltonian describing the one-electron motions may be written as $H = -\frac{\hbar^2}{2m}\nabla^2 + V_A + V_B + V_{ee}$, where $V_A$, $V_B$ represents the electron-nuclei interactions due to A and B atoms, respectively; $V_{ee}$ is the electron-electron interaction. Specially, the atomic levels $\varepsilon_{nA} \approx \langle \phi_{nA} | H | \phi_{nA} \rangle$, $\varepsilon_{nB} \approx \langle \phi_{nB} | H | \phi_{nB} \rangle$. The eigen-energy due to the basis $\psi_n$ may be evaluated as follows: $\varepsilon_n = \langle \psi_n | H | \psi_n \rangle$. By considering terms due to the nearest neighboring interactions only, one has $\langle \phi_{nAi} | H | \phi_{nAj} \rangle \approx \delta_{ij} \varepsilon_{nA}$, $\langle \phi_{nBi} | H | \phi_{nBj} \rangle \approx \delta_{ij} \varepsilon_{nB}$, and the overlap integral is given by $\langle \phi_{nAi} | H | \phi_{nBj} \rangle = \langle \phi_{nBj} | H | \phi_{nAi} \rangle \equiv -V_{AB} < 0$. It follows that

$$\varepsilon_n = \sum_{i=1}^{n_A} |g_{Ai}|^2 \varepsilon_{nA} + \sum_{j=1}^{n_B} |g_{Bj}|^2 \varepsilon_{nB} + 2(\sum_{i,j} Re[g_{Ai} g_{Bj}^*]) \langle \phi_{nAi} | H | \phi_{nBj} \rangle =$$

$$\sum_{i=1}^{n_A} |g_{Ai}|^2 \varepsilon_{nA} + \sum_{j=1}^{n_B} |g_{Bj}|^2 \varepsilon_{nB} - 2V_{AB} \sum_{i,j} Re[g_{Ai} g_{Bj}^*]. \tag{A3}$$

Compared to the single atomic states, it is clear that the energy gain upon the formation of A-B bonds comes mainly from the last term, which is due to overlap of wave functions at neighboring atomic sites. Using the constrain condition $\sum_{i=1}^{n_A} |g_{Ai}|^2 + \sum_{j=1}^{n_B} |g_{Bj}|^2 = 1$, the largest energy gain due to overlap integral is reached when $|g_{Ai}|^2 = |g_{Bj}|^2 = \frac{1}{N}$. Consequently, the largest energy gain due to the overlap of the total wave function $\Psi$ is given by the expanding coefficients which satisfy $|c_{An}|^2 = |c_{Bn}|^2 = \frac{|c_n|^2}{N}$. This corresponds to a uniformly distribution at each atomic site with a weight of $|c_n|^2$ for the basis $\psi_n$.